\documentclass[a4paper]{article}

\usepackage{INTERSPEECH2021}

\usepackage{url}
\usepackage{graphicx}
\usepackage{amsmath}
\usepackage{hyperref}
\usepackage{booktabs}
\usepackage{array}

\newcolumntype{P}[1]{>{\centering\arraybackslash}p{#1}}
\newcolumntype{M}[1]{>{\centering\arraybackslash}m{#1}}
\usepackage{bm}
\usepackage{amssymb}
\usepackage{amsmath}
\usepackage{color}
\usepackage{multirow}
\usepackage{tablefootnote}

\usepackage{xcolor}

\providecommand{\fj}[1]{\textcolor{blue}{{#1}}}

\title{An Empirical Study on Channel Effects for Synthetic Voice Spoofing Countermeasure Systems}
\name{You Zhang$^1$, Ge Zhu$^1$, Fei Jiang$^{1,2}$, Zhiyao Duan$^1$}

\address{
  $^1$University of Rochester, Rochester, NY, USA \\
  $^2$Beijing Institute of Technology, Beijing, China}
\email{\{you.zhang, ge.zhu, fei.jiang, zhiyao.duan\}@rochester.edu \\
\href{https://github.com/yzyouzhang/Empirical-Channel-CM}{\fj{\texttt{https://github.com/yzyouzhang/Empirical-Channel-CM}}}}

\begin{document}

\maketitle

\begin{abstract}
Spoofing countermeasure (CM) systems are critical in speaker verification; they aim to discern spoofing attacks from bona fide speech trials. 
In practice, however, acoustic condition variability in speech utterances may significantly degrade the performance of CM systems.
In this paper, we conduct a cross-dataset study on several state-of-the-art CM systems and observe significant performance degradation compared with their single-dataset performance. 
Observing differences of average magnitude spectra of bona fide utterances across the datasets, we hypothesize that channel mismatch among these datasets is one important reason. We then verify it by demonstrating a similar degradation of CM systems trained on original but evaluated on channel-shifted data.
Finally, we propose several channel robust strategies (data augmentation, multi-task learning, adversarial learning) for CM systems, and observe a significant performance improvement on cross-dataset experiments. 


\end{abstract}


\section{Introduction}
Automatic speaker verification (ASV) systems are vulnerable to spoofing attacks, where attackers pretend to be the target speaker by presenting false but similar-to-bona-fide speech trials~\cite{evans2014speaker}. Spoofing countermeasure (CM) systems aim to detect such attacks. Spoofing attacks are considered physical access (PA) if they are utterances presented to the microphone of the ASV system, and logical access (LA) if they bypass the microphone and feed to the verification algorithm directly. In a common anti-spoofing setup as in ASVspoof2019~\cite{todisco2019asvspoof}, the PA scenario features replay attacks using various playback devices, while the LA scenario features synthetic utterances generated by text-to-speech (TTS) and voice conversion (VC) algorithms. 

Recently, deep learning technologies have shown great success in learning discriminative speaker embeddings to classify spoofing attacks from bona fide speech in the LA scenario~\cite{nautsch2021asvspoof}. The CM community has been exploring the usage of different input speech features~\cite{das2019long, yang2019significance, yang2020long, tak2020end}, model architectures~\cite{lai2019assert, lavrentyeva2019stc, monteiro2020generalized, parasu2020investigating, li2020replay}, and loss functions~\cite{gomez2020kernel, chen2020generalization, zhang2020one} to improve the performance on detecting synthetic attacks.


However, several cross-dataset studies~\cite{korshunov2016cross, paul2017generalization, himawan2019deep, das2020assessing} in anti-spoofing show significant performance degradation from single-dataset studies. 
For example, when systems are trained on LA but tested on PA, performance degradation happens, and suggested solutions include the usage of more generalized speech features~\cite{paul2017generalization, das2020assessing} and domain adaptation~\cite{himawan2019deep}. Another more surprising performance degradation happens when a state-of-the-art CM system is trained and tested on different LA datasets~\cite{das2020predictions}. The authors suggested that it is because some unseen attacks are more challenging. While we agree that this can be an important reason, we also think that other differences, such as channel variation across datasets could be possible reasons.
This motivates us to systematically conduct a cross-dataset study on synthetic voice spoofing CM systems. 


Such cross-dataset studies are important in the design of robust CM systems. Due to limited access of training data and its channel variation, CM systems may be frail in practice. Here \textit{channel effects} refer to audio effects imposed onto the speech signal throughout the entire recording and transmission process, including reverberation of recording environments, frequency responses of recording devices, and compression algorithms in telecommunication.
Without properly considering and compensating for these effects, CM systems may overfit to the limited channel effects presented in the training set and fail to generalize to unseen channel variation. This issue has been studied in replay attacks~\cite{wang2019cross}, but little attention is paid in the LA scenario.
For example, the bona fide speech utterances in ASVspoof2019LA were all from the VCTK corpus, and the TTS/VC systems used in generating LA attacks were all trained on the VCTK corpus. This may introduce strong biases to CM systems on the limited channel variation.  

In this work, we first conduct a cross-dataset study of three state-of-the-art CM systems between ASVspoof2019LA, ASVspoof2015, and VCC2020. Observing significant performance degradation on all CM systems, we hypothesize that the channel effect mismatch between these datasets is one important reason for the degradation. To test our hypothesis, we first compare the average magnitude spectra across all bona fide utterances among these three datasets and observe significant mismatches. We then conduct a controlled cross-channel experiment by training the three CM systems on ASVspoof2019LA and evaluating them on the evaluation set of its channel-augmented version, ASVspoof2019LA-Sim, which is generated by passing ASVspoof2019LA utterances through an acoustic simulator~\cite{ferras2016large}; our hypothesis is again verified by consistent performance degradation across the three CM systems. 
Finally, we propose several strategies to improve channel robustness leveraging the channel-augmented data. Results show that these strategies successfully improve the cross-dataset performance of all three CM systems.

As we conduct this study, we notice that the LA sub-challenge of ASVspoof 2021~\cite{asvcmwebsite} also intends to consider channel robustness in its evaluation mechanism. We believe that our study will provide useful insights into this research direction.






\vspace{-8pt}

\section{Cross-Dataset Studies}
\label{sec:case_study}
In this section, we take three state-of-the-art CM systems and three commonly used anti-spoofing datasets to extensively study the performance degradation issue in cross-dataset evaluation for synthetic voice spoofing CM systems. 


\subsection{Datasets}
We employ three datasets containing both bona fide speech and synthetic speech generated by TTS or VC algorithms. 


\textbf{ASVspoof2019LA}~\cite{wang2020asvspoof} is a large-scale dataset used in the LA sub-challenge of ASVspoof2019. It contains a large variety of up-to-date TTS and VC algorithms forming a diverse collection of attacks. The bona fide speech was collected from the VCTK corpus~\cite{yamagishi2019cstr}. Training and Development sets share the same attacks (A01-A06), but the evaluation set contains totally different attacks (A07-A19). 

\textbf{ASVspoof2015}~\cite{wu2015asvspoof} is the database for the 2015 edition of the ASVspoof challenge, which only deals with synthetic voice spoofing attacks. 
The training set includes S01-S05 attacks and the evaluation set includes S01-S10, which have both known and unknown attacks.

\textbf{VCC2020} (Voice Conversion Challenge 2020)~\cite{yi2020voice} distributed a new dataset that aims to develop systems for converting speech from a source speaker to a target speaker. The participating teams submitted their converted speech developed on the training data. The utterances in the training data provided by the organizers are considered bona fide trials, while the converted utterances generated by each submitted VC system are considered spoofing attacks. Different from the previous two datasets, VCC2020 is multilingual.


\subsection{Experimental Setup}
\label{ssec:exp_set}

We select three state-of-the-art CM systems that show top performance on the ASVspoof2019LA database: LCNN~\cite{lavrentyeva2019stc}, ResNet~\cite{monteiro2020generalized}, and ResNet-OC~\cite{zhang2020one}.
These CM systems are trained on the training set of ASVspoof2019LA~\cite{wang2020asvspoof}, and validated on its development set. For evaluation, we use only the evaluation sets of ASVspoof2019LA and ASVspoof2015, and the complete VCC2020. In this way, all spoofing attacks in evaluation are unknown.

For all of the CM systems, the 60-d linear-frequency cepstral coefficients (LFCC) are extracted as speech features from each frame of the utterances. The frame length is 20ms and the hop size is 10ms. To form batches, we set 750 frames as the fixed length; We use repeat padding for shorter trials, and we randomly choose a consecutive  750-frame segment for longer trials. The learning rate is initially set to 0.0003 with 50\% decay for every 10 epochs. We train the network for 100 epochs on a single NVIDIA GTX 1080 Ti GPU. Finally, we select the model with the lowest validation loss for evaluation.

Each CM system outputs a score to indicate the confidence that the given utterance is bona fide. 
Equal Error Rate (EER) is calculated by setting a threshold on the CM score such that the false alarm rate is equal to the miss rate. We use EER since it is used across all ASVspoof challenge series.

\subsection{Results and Analyses}
\label{ssec:pre_result}

In Table~\ref{tab:cross}, we demonstrate EER degradation across datasets for all three CM systems. As a sanity check, the result for LCNN is consistent with that in~\cite{das2020assessing} on ASVspoof2015 and ~\cite{das2020predictions} on VCC2020. For the sake of space and without loss of generality, here we only analyze the ResNet-OC CM system, as it achieves the lowest EER among all single-system CM on the evaluation set of the ASVspoof2019LA dataset~\cite{zhang2020one}.

\begin{table}[htbp]
\caption{EER performance across different evaluation datasets (ASVspoof2019LA-eval, ASVspoof2015-eval, VCC2020). All of the three CM systems are trained on the training set of ASVspoof2019LA and validated on its development set.}
\renewcommand{\arraystretch}{1.0}
\small
\centering
\scalebox{0.85}{
\begin{tabular}{c|ccc}
\hline \hline
EER (\%)        & \multicolumn{3}{c}{CM Systems}           \\
Evaluation Datasets             & LCNN~\cite{lavrentyeva2019stc}  & ResNet~\cite{monteiro2020generalized} & ResNet-OC~\cite{zhang2020one}\\
\hline
2019LA-eval      &          3.25      &   5.23    &      2.29  \\
2015-eval     &   24.55          &     37.11 &  26.30    \\
VCC2020 &   33.78               &  36.09   &     41.66      \\
\hline\hline
\end{tabular}
}
\label{tab:cross}
\end{table}

\begin{figure}[htbp]
  \centering
  \includegraphics[width=\linewidth]{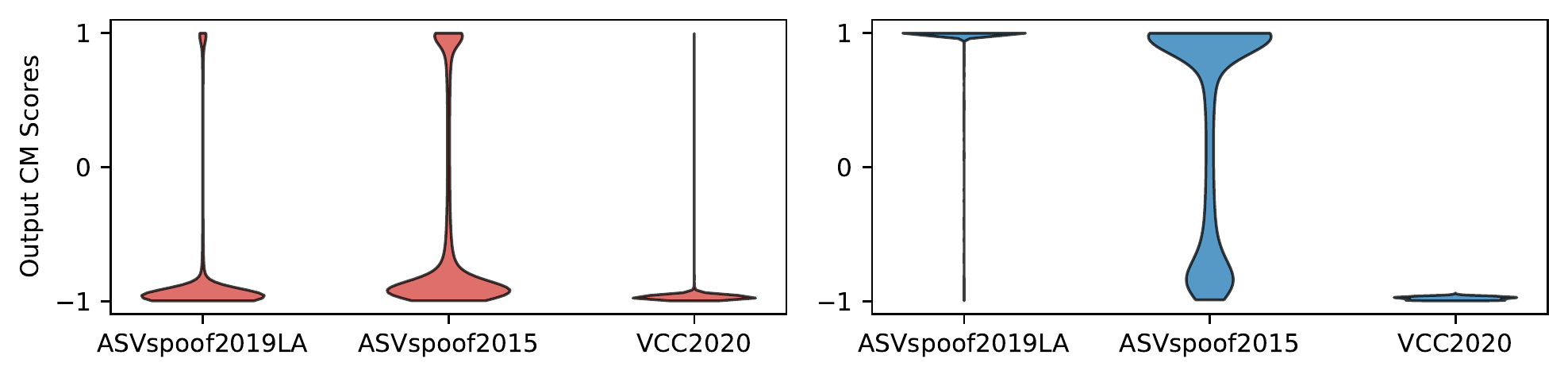}
  \caption{Score distributions of ResNet-OC method on  spoofing attacks (left) and bona fide (right) of cross-dataset evaluation.
  }
  \label{fig:score_dist}
\end{figure}

The EER degradation could be because that the score distribution of spoof trials shifts up, or that of bona fide trials shifts down. We hence plot the score distributions of ResNet-OC in Figure~\ref{fig:score_dist}.
We observe that the unknown spoofing attacks are mostly correctly scored across the three datasets, verifying that ResNet-OC is capable of detecting unseen attacks. However, the score distribution of bona fide trials shifts down significantly from ASVspoof2019LA to ASVspoof2015 and VCC2020, which would cause many false alarm errors. This suggests that the main cause of the EER degradation is some differences in bona fide speech, among which, channel variation 
is worth checking. 

\begin{figure}[!htbp]
  \centering
  \includegraphics[width=\linewidth]{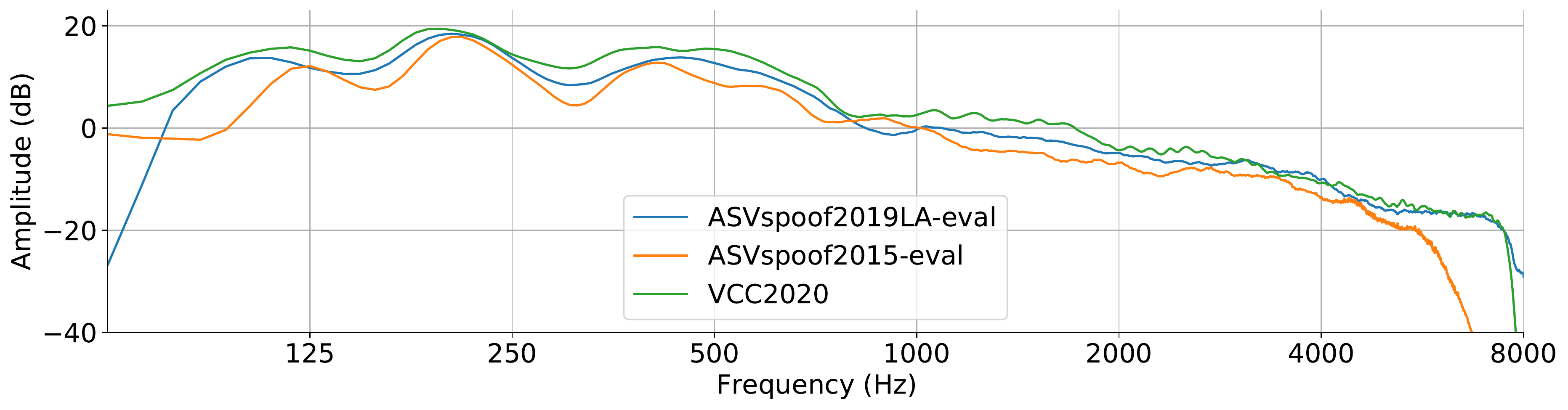}
  \caption{Average magnitude spectra of bona fide utterances across different datasets. 
}
  \label{fig:eq_curve_cross}
\end{figure}

In Figure~\ref{fig:eq_curve_cross}, we show the average magnitude spectrum across all bona fide utterances of each dataset as an indication of its channel effect. We can see that the average spectra are very different among the three datasets
. 
With this observation, we hypothesize that channel mismatch is an important reason for the EER degradation, and will test this hypothesis through a controlled experiment next.

\subsection{Channel Simulation}
\label{ssec:chan_sim}

\begin{figure}[h]
  \centering
  \includegraphics[width=\linewidth]{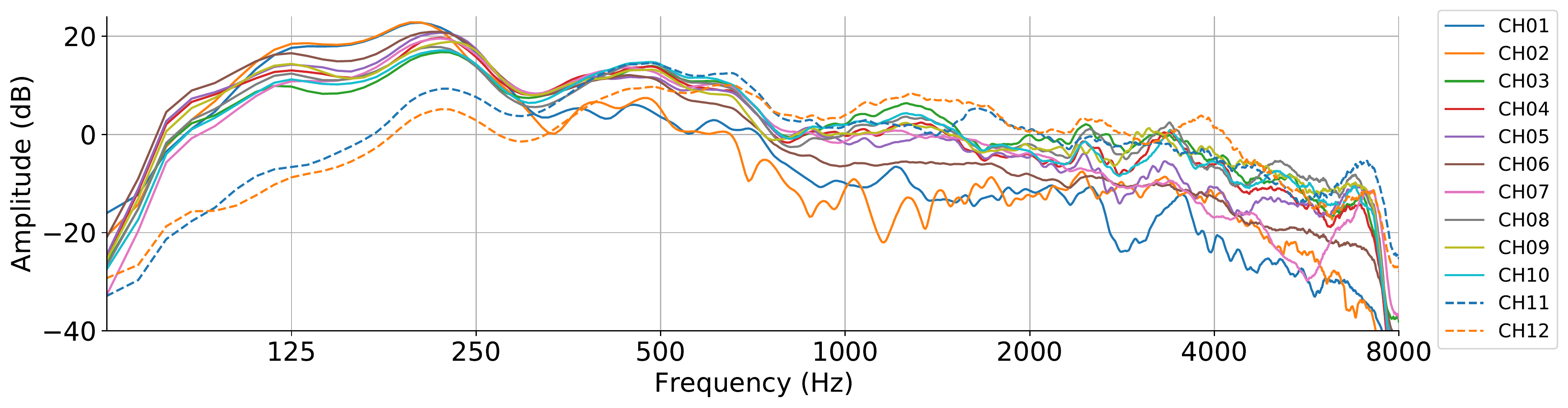}
  \caption{Average magnitude spectra of channel-shifted bona fide utterances in the evaluation set of \textit{ASVspoof2019LA-Sim} using different channel IRs.}
  \label{fig:eq_curve_device}
\end{figure}

We augment the channel effects of the ASVspoof2019LA dataset through an open-source channel simulator~\cite{ferras2016large}.
This channel simulator provides three types of degradation processes including additive noise, telephone and audio codecs, and device/room impulse responses (IRs). In our simulation, we choose 12 out of 74 different device IRs and apply each of them to all utterances of ASVspoof2019LA. This channel-augmented dataset is named \textbf{ASVspoof2019LA-Sim}, and the same train/dev/eval split is followed from ASVspoof2019LA. We do not use additive noise, audio codecs, or reverberation IRs in our simulation. 
The average magnitude spectra of the augmented bona fide utterances in the ASVspoof2019LA-Sim evaluation set using each device IR are plotted in Figure~\ref{fig:eq_curve_device}. 

We then test the performance of the three CM systems selected in Section~\ref{ssec:exp_set} on the evaluation set of ASVspoof2019LA-Sim. As a reminder, all three CM systems are trained and validated on the training and development sets of ASVspoof2019LA (i.e., without augmented channel-shifts). Table~\ref{tab:channel} shows the results. The average and standard deviation of EERs across all of the 12 simulated channels are calculated. We observe that the average EER drops significantly from ASVspoof2019LA-eval and the standard deviation is quite large. We conclude that the channel mismatch between training and evaluation is indeed an important reason for the performance degradation. 

\begin{table}[htbp]
\caption{EER performance on ASVspoof2019LA-Sim-eval. Average and standard deviation EERs are calculated across the 12 simulated channels. All of the three CM systems are trained on ASVspoof2019LA-train.}
\renewcommand{\arraystretch}{1}
\small
\centering
\scalebox{0.85}{
\begin{tabular}{c|ccc}
\hline \hline
   EER (\%)  & \multicolumn{3}{c}{CM Systems}           \\
  Statistics   & LCNN~\cite{lavrentyeva2019stc}  & ResNet~\cite{monteiro2020generalized} & ResNet-OC~\cite{zhang2020one}\\
\hline
Avg. (CH01-CH12)    &  27.75    &   48.78      &   40.46  \\
Std. ~(CH01-CH12)  &   7.44      &   18.80     &   11.22 \\
\hline\hline
\end{tabular}
}
\label{tab:channel}
\end{table}





\section{Channel Robust Strategies}
We propose several strategies to improve the channel robustness of CM systems using the channel-shifted data. Specifically, we create a \textit{channel-augmented training set} containing the original ASVspoof2019LA training data and only 10 out of the 12 channel shifts of the ASVspoof2019LA-Sim training data. For evaluation, we use all 12 channel shifts of the ASVspoof2019LA-Sim evaluation data and their original utterances.



\begin{figure}[]
  \centering
  \includegraphics[width=0.92\linewidth]{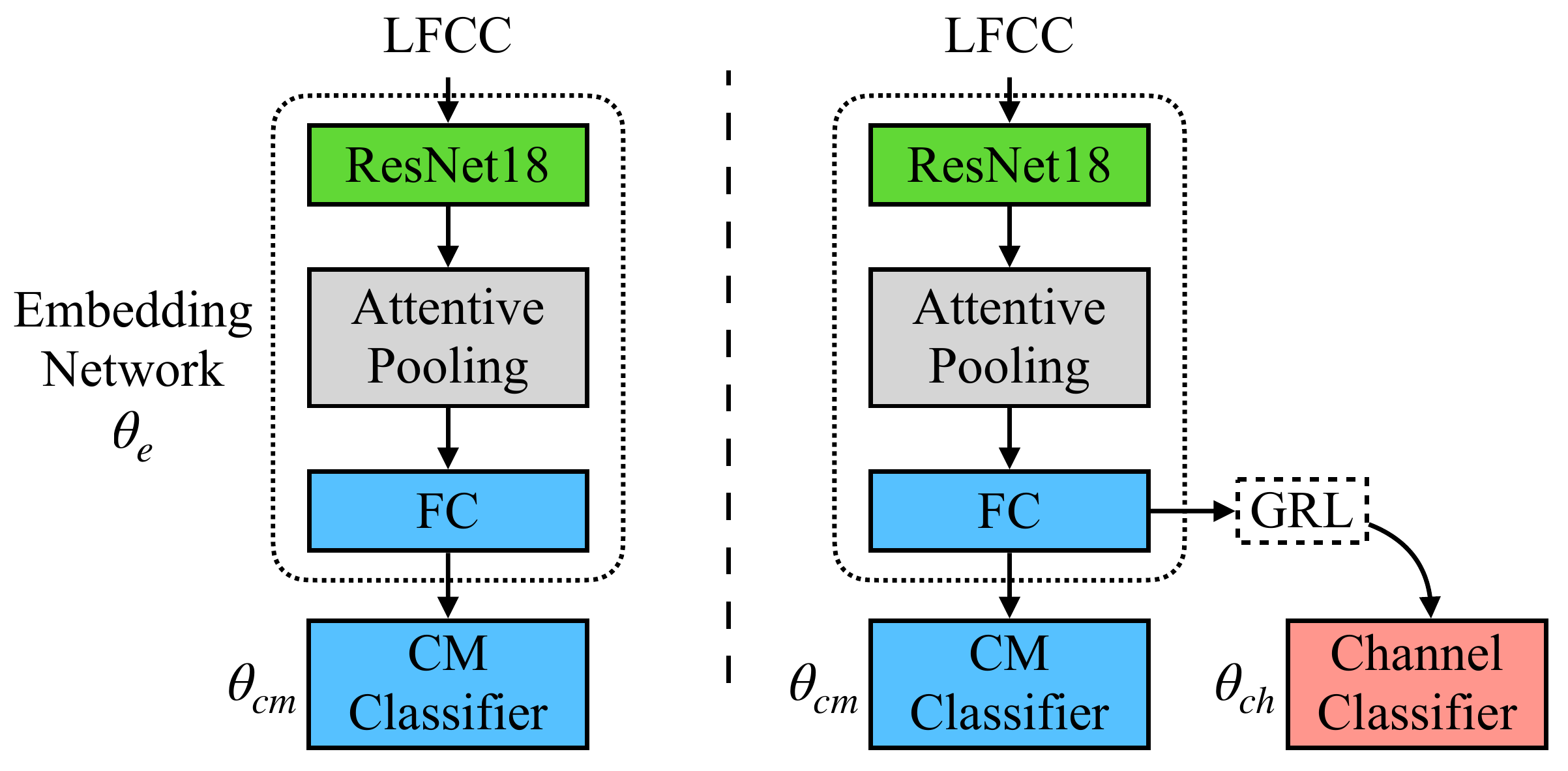}
  \caption{Model structure of the proposed channel robust strategies. Left: Vanilla model and AUG. Right: MT-AUG (w/o GRL) and ADV-AUG (w/ GRL).} 
  \label{fig:model_struc}
\end{figure}

Similar to Section~\ref{ssec:pre_result}, without loss of generality and for the sake of space, we only demonstrate these strategies on the ResNet-OC CM system~\cite{zhang2020one}.
Its model architecture is illustrated on the left side of Figure~\ref{fig:model_struc}. An embedding network based on ResNet18 with attentive pooling, parameterized by $\theta_e$, aims to learn a discriminative speech embedding, which is then classified by another fully connected (FC) layer, parameterized by $\theta_\textit{cm}$, into spoofing attacks or bona fide speech. The loss function is OC-Softmax. Without the channel robust strategies, ResNet-OC is trained only on the ASVspoof2019LA training set, and this model is named the \textbf{Vanilla} model as a baseline.



\subsection{Proposed Strategies}

Utilizing utterances from the ASVspoof2019LA-Sim training set, we proposed three channel-robust strategies.

\textbf{Augmentation (AUG)} uses the same model architecture as the Vanilla model (Figure~\ref{fig:model_struc} left side) but is trained on the channel augmented training set mentioned above.

\textbf{Multi-Task Augmentation (MT-AUG)} adds a channel classifier, parameterized by $\theta_\textit{ch}$, to the vanilla model architecture to form a multi-task learning setup. The overall model structure is shown on the right side of Figure~\ref{fig:model_struc}, skipping the ``GRL'' module. This channel classifier uses two fully connected layers to map deep speech embeddings to the channel labels, and uses the cross entropy loss. The overall training objective of MT-AUG is thus:
\begin{equation}
\begin{aligned}
(\hat{\theta}_{e}, \hat{\theta}_\textit{cm}, \hat{\theta}_\textit{ch}) &=\underset{\theta_{e}, \theta_\textit{cm}, \theta_\textit{ch}}{\arg \min } \mathcal{L}_{\textit{cm}}\left(\theta_{e}, \theta_\textit{cm}\right)+\lambda \mathcal{L}_{\textit {ch}}\left(\theta_{e}, \theta_\textit{ch}\right).
\end{aligned}
\end{equation}


\textbf{Adversarial Augmentation (ADV-AUG)} inserts a Gradient Reversal Layer (GRL)~\cite{ganin2015unsupervised} between the embedding network and the channel classifier. The loss of the channel classifier is now backpropagated through the GRL, with the sign reversed, to the embedding network. Therefore, the embedding network aims to maximize the channel classification error while the channel classifier aims to minimize it, forming an adversarial training paradigm.
When equilibrium is reached, the learned speech embeddings would be channel-agnostic, making the CM classifier robust to channel variation. 
\begin{equation}
\begin{aligned}
(\hat{\theta}_{e}, \hat{\theta}_\textit{cm}) &=\underset{\theta_{e}, \theta_\textit{cm}}{\arg \min } \mathcal{L}_{\textit{cm}}\left(\theta_{e}, \theta_\textit{cm}\right)-\lambda \mathcal{L}_{\textit{ch}}(\theta_{e}, \hat{\theta}_\textit{ch}) \\
(\hat{\theta}_\textit{ch}) &=\underset{\theta_\textit{ch}}{\arg \min } \mathcal{L}_{\textit{ch}}(\hat{\theta}_{e}, \theta_\textit{ch})
\end{aligned}
\end{equation}

It is noted that GRL has been employed in various speech processing tasks such as phonetic-informed speaker embedding~\cite{wang2019usage}, channel-invariant speaker verification~\cite{meng2019adversarial, chen2020channel}, speaker-invariant speech emotion recognition~\cite{li2020speaker, yin2020speaker}, and cross-domain replay spoofing attack detection~\cite{wang2019cross}.











\subsection{In-Domain Test}
In this test, we evaluate the proposed strategies on the evaluation set of our simulated ASVspoof2019LA-Sim dataset. Out of the 12 channel shifts in the evaluation set, 10 have also been used in forming the channel-augmented training set, with which we trained the above strategies. Therefore, our test results contain both seen (CH01-10) and unseen (CH11-12) channel shifts.

Table~\ref{tab:method_comp} shows EER statistics. We can see that the average EER on seen channels and the EERs on unseen channels of all three training strategies decrease much from the vanilla model. 
This shows that the proposed strategies do improve the channel robustness of the CM system.
The significant decrease of the standard deviation of EER across the 10 seen channels also suggests that the strategies make the CM system less sensitive to channel variation. 
Comparing the three strategies in terms of EER, ADV-AUG performs the best on seen channels, while AUG performs the best on unseen channels.

\begin{table}[htbp]
\caption{EER performance comparison of the proposed strategies and the vanilla model on ASVspoof2019LA-Sim-eval. The proposed strategies are trained on the augmented training set.}
\centering
\scalebox{0.85}{
\begin{tabular}{c|c|ccc}
\hline \hline
   EER (\%)                & \multicolumn{4}{c}{Methods}           \\
  & Vanilla & AUG & MT-AUG & ADV-AUG \\
\hline
Avg. (CH01-10)      &  38.14    &  4.43       &  4.29   & 3.92 \\
Std. ~(CH01-10)   &   10.83       &    0.75    &    0.46  & 0.43\\
\hline
CH11 & 54.98 & 3.58  & 4.59  &3.78 \\
CH12 & 49.17 & 4.41  & 7.08   &6.28 \\
\hline\hline
\end{tabular}
}
\label{tab:method_comp}
\end{table}

\begin{figure}[htbp]
  \centering
  \includegraphics[width=\linewidth]{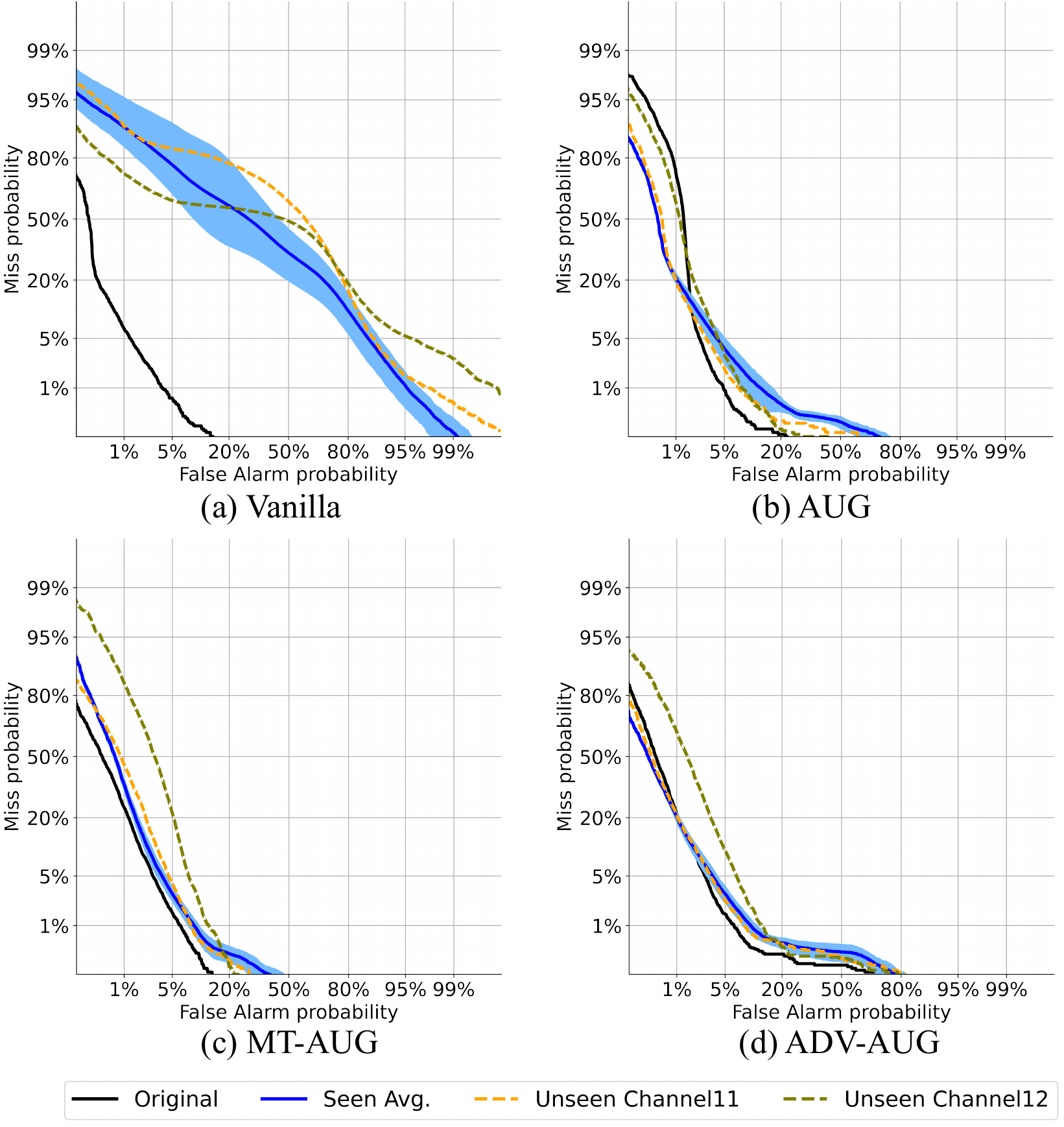}
  \caption{DET curves of the vanilla model and the proposed channel robust strategies, evaluated on the original ASVspoof2019LA-eval and the simulated ASVspoof2019LA-Sim-eval (with 12 channel effects (10 seen, 2 unseen))}
  \label{fig:det_compare}
\end{figure}

The detection error tradeoff (DET) curve is often used to show the tradeoff between miss and false alarm errors in detection tasks~\cite{martin1997det}.
We compute a DET curve for the original and each channel-shifted data subset 
. In Figure~\ref{fig:det_compare}, we plot the DET curve on the original ASVspoof2019LA-eval dataset, the average of the DET curves across all seen channels (CH01-CH10) with its one standard deviation (SD) above and below, and the DET curves of the two unknown channels (CH11, CH12). All of the curves are shown in the normal deviate scale.



For channel-robust CM systems, the SD region of DET curves of seen channels is expected to be narrow, and the curves of the original data and unknown channels are expected to be within or close to the SD region. From Figure~\ref{fig:det_compare}, the vanilla system does not show such properties. The proposed approaches, in contrast, all show much narrower SD regions, and the DET curves for the original data and unseen channels are also much closer to the SD region. This again suggests that the proposed approaches improve the channel robustness of the CM system.

\subsection{Out-of-Domain (Cross-Dataset) Test}
We perform a cross-dataset evaluation as in Section~\ref{sec:case_study}, and the EER results are shown in Table~\ref{tab:method_cross}. Compared to the Vanilla model, our proposed channel-robust strategies show slight degradation on the in-domain dataset, ASVspoof19LA-eval. However, they show significant improvement on both out-of-domain datasets, ASVspoof2015-eval and VCC2020, verifying our hypothesis of channel mismatch among these datasets and the effectiveness of the proposed strategies.

\begin{table}[htbp]
\caption{EER comparison of the proposed strategies and the vanilla model on cross-dataset evaluation.}
\renewcommand{\arraystretch}{1.0}
\centering
\scalebox{0.85}{
\begin{tabular}{c|c|ccc}
\hline \hline
     EER(\%)             & \multicolumn{4}{c}{Methods}     
                  \\
Evaluation Datasets             & Vanilla & AUG & MT-AUG & ADV-AUG \\
\hline
2019LA-eval   &   2.29   &    2.92     &  3.41   &  3.23\\
2015-eval  &   26.30    &   16.25     &  22.10  & 14.38 \\
VCC2020    &  41.66   &   30.51      &   28.85  &  27.07\\
\hline\hline
\end{tabular}
}
\label{tab:method_cross}
\end{table}

\begin{figure}[htbp]
  \centering
  \includegraphics[width=\linewidth]{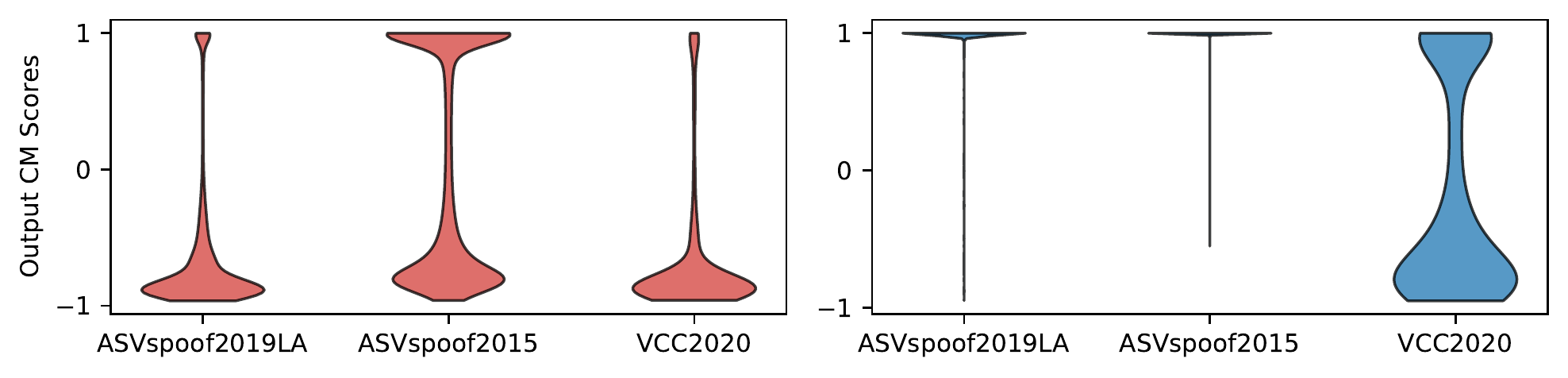}
  \caption{Score distributions of ADV-AUG strategy on  spoofing attacks (left) and bona fide (right) of cross-dataset evaluation.}
  \label{fig:new_score_dist}
\end{figure}

As ADV-AUG achieves the best cross-dataset performance among all channel robust strategies, we show its new score distribution in Fig.~\ref{fig:new_score_dist}. Compared with Fig.~\ref{fig:score_dist}, the score distributions of spoofing attacks are not changed much, 
while those of bona fide trials in ASVspoof2015 and VCC2020 are all moved up. This suggests that the channel mismatch issue mainly resides on bona fide speech, and the proposed strategies are useful. On the other hand, this move-up is not sufficient for VCC2020, suggesting the limitation of the proposed strategies. The remaining performance gap may also be due to other unique properties of VCC2020 such as its multilingual and more challenging attacks. 


\section{Conclusions}
In this paper, we observed significant performance degradation of several state-of-the-art CM systems when they are trained on ASVspoof2019LA and tested on ASVspoof2015 and VCC2020.
We then hypothesized that the channel effect is one reason for the performance degradation, after observing that the average magnitude spectrum of bona fide speech is different across the datasets.
We further verified this hypothesis by testing a CM system on a channel-shifted dataset using various device IRs and observing a similar performance degradation. 
We then proposed several strategies 
to improve the robustness of CM systems to channel variation, and obtained significant improvement in both in-domain and cross-dataset tests. 
For future work, we plan to investigate other potential factors that may cause cross-dataset performance degradation of CM systems.



\section{Acknowledgements}
This work was supported by National Science Foundation grant No. 1741472 and funding from Voice Biometrics Group. The authors would also like to thank Dr. Xin Wang from National Institute of Informatics, Tokyo, Japan for valuable discussions.

\vfill\pagebreak

\bibliographystyle{IEEEtran}

\bibliography{mybib}


\end{document}